\documentclass[nojss,nofooter]{jss}
\usepackage{thumbpdf,lmodern}
\usepackage{booktabs}
\usepackage{multirow}
\usepackage{amsmath,amssymb}
\usepackage{adjustbox}
\usepackage{framed}
\usepackage{float}

\newtheorem{Def}{Definition}

\newtheorem{eje}{Example}
\newtheorem{teo}{Theorem}

\author{Izhar Asael Alonzo Matamoros \\ Universidad de Cantabria \\izhar.alonzo@unah.edu.hn \And 
	Alicia Nieto-Reyes \\ Universidad de Cantabria \\alicia.nieto@unican.es}
\Plainauthor{ }
\title{An \proglang{R} Package for Normality in Stationary Processes}
\Plaintitle{An R Package for Normality in Stationary Processes}
\Shorttitle{An \proglang{R} package for normality in stationary processes}
\Abstract{
 Normality is the main assumption for analyzing dependent data in several time series models, and tests of normality have been widely studied in the literature, however, the implementations of these tests are limited. The \pkg{nortsTest} package performs the tests of \textit{Lobato and Velasco, Epps, Psaradakis and Vavra} and \textit{random projection} for normality of stationary processes. In addition, the package offers visual diagnostics for checking  stationarity and normality assumptions for the most used time series models in  several \proglang{R} packages. The aim of this work is to show the functionality of the package, presenting each test performance with simulated examples, and the package utility for model diagnostic in time series analysis.
}
\Plainkeywords{Gaussian process, hypothesis test, stochastic process}
\Address{
	Izhar Asael Alonzo Matamoros\\
	Universidad Nacional Autónoma de Honduras\\
	Facultad de Ciencias, Escuela de Matemática\\
	Blvd. Suyapa, UNAH, edificio F1\\
	Francisco Morazán, Honduras\\
	E-mail: \email{izhar.alonzo@unah.edu.hn}\\
	URL: \url{https://github.com/asael697}\\
	
	Alicia Nieto Reyes\\
	Department of Mathematics, Statistics and Computer Science \\
	University of Cantabria\\
	Avd. de los Castros s/n. 39005
	 Santander, Spain\\
	E-mail: \email{alicia.nieto@unican.es}\\
}
\begin{document}
%
\addcontentsline{toc}{section}{Introduction}
\section{Introduction}\label{sec:intro}
Normality (\textit{a set of observations being sampled from a Gaussian process}) is an important assumption in a wide variety of statistical models. Therefore, developing procedures for testing this assumption is a topic that has gained popularity over several years. Most of the existing  literature, and implementation, is dedicated to independent and identically distributed random variables  \citep{Dagostino1987}; and  there are no results showing that these tests are consistent in the context of stationary processes.  For this context, a small number of tests have been proposed over the years, but, as far as we know, there exist no \proglang{R} package or consistent implementation of them.

The proposed \pkg{nortsTest} package provides four test implementations for normality in stationary processes. The aim of this work is to present a review of these tests and introduce the  package functionality. The implemented tests are: (i) the \textit{Epps} test \citep{epps1987} based on the characteristic function, (ii) the corrected \textit{Skewness-Kurtosis} (SK) test implemented by  \citet{Lobato2004}, (iii) the \textit{random projection test} proposed by \cite{nietoreyes2014} and (iv) the \textit{Psadarakis and V\'avra test} \citep{vavra2017} that uses a bootstrap approximation of the \cite{anderson1952} test statistic for stationary linear processes. Additionally, we propose  the \code{check_residual()} function for checking the assumptions in time-series models, which returns a report of tests for stationarity, seasonality, and normality as well as diagnostic plots for visual check. This function supports models from the most used packages for time-series analysis, such as the  packages \pkg{forecast} \citep{Rob2007} and \pkg{aTSA} \citep{aTSA}, and even functions in the base \proglang{R} \citep{R}; for instance, it supports the  \code{HoltWinters} (\pkg{stats} \proglang{R} package) function for the Holt and Winters method \citep{Holt2004}.

Section \ref{sec:concepts} provides the theoretical background, including preliminary concepts and results. Section \ref{sec:normality} introduces the normality tests for stationary processes, each subsection introducing a test framework and including examples of the tests functions with simulated data. Section \ref{sec:simulation} provides numerical experiments with simulated data and a real data application. In Subsection \ref{sub:numerical} reports a simulation study for all the implemented tests, and Subsection \ref{sub:application} shows the functionality of the package  for model checking in a real data application:  the \textit{carbon dioxide} data measured in the Malua Loa Obsevatory \citep{astsa} is analyzed using a state space model from the \pkg{forecast} package and the model assumptions are evaluated using the proposed \code{check_residuals()} function. Section \ref{sec:conclusion} discusses the package functionality and provides our conclusions. Furthermore, we include our future work on the package.
%
%
\addcontentsline{toc}{section}{Preliminary Concepts}
\section{Preliminary concepts} \label{sec:concepts}
This section provides some theoretical aspects of stochastic processes that are a necessary theoretical framework for the following sections. The presented definitions and results can be found in \cite{shumway2010} and \cite{Ts2010}.

For the purpose of this work,  $T$ is a set of real values denoted as time, $T \subseteq \mathbb{R},$ for instance $T=\mathbb{N}$ or $T=\mathbb{Z},$ the naturals or integer numbers respectively. We denote by $X:=\{X_t\}_{t\in T}$ a \textit{stochastic process} with $X_t$ a real random variable for each $t\in T.$ Following this notation, a  \textit{time-series} is just a finite collection of ordered observations of $X$ \citep{shumway2010}. An important measure for a stochastic process is its mean function $\mu(t) := E[X_t]$ for each $t \in T$, where $E[\cdot]$ denotes the usual expected value of a random variable. A generalization of this measure is the k-th order centered moment function $\mu_k(t) := E[(X_t -\mu(t))^k]$ for each $t \in T$ and $k > 1;$ with the process variance function the second order centered moment, $\sigma^2(t) := \mu_2(t)$. Other important indicators are the auto-covariance and auto-correlation functions, which measure the linear dependency between two different time points of a given process. For any $t,s \in T,$ they are, respectively, 
$$
\gamma(t,s) := E[(X_t -\mu(t))(X_s - \mu(s))] \mbox{ and } \rho(t,s) := \dfrac{\gamma(t,s)}{\sqrt{\mu_2(t)}\sqrt{\mu_2(s)}}.$$
Other widely used indicator functions  for the analysis of  processes are the skewness and kurtosis functions, defined for each $t\in T$ as $s(t) := \mu_3(t)/[\mu_2(t)]^{3/2}$ and $k(t) := \mu_4(t)/[\mu_2(t)]^2$ respectively. 

A generally used assumption for stochastic processes is stationarity. It has a key role in forecasting procedures of classic time-series modeling \citep{Ts2010} or as a principal assumption in de-noising methods for signal theory \citep{W2006}. 
\begin{Def}
  A stochastic process $X$ is said to be \emph{strictly stationary} if, for every collection $\tau = \{t_1,t_2,\ldots, t_k\} \subset T$ and $h > 0$, the joint distribution of $\{X_t\}_{t \in \tau}$ is identical to that of $\{X_{t+h}\}_{t \in \tau}.$
\end{Def}  
The previous definition is strong for applications. A milder version of it, which makes use of the process first two moments, is weak stationarity.
\begin{Def}
  A stochastic process $X$ is said to be  \emph{weakly stationary} if its mean function is constant in time, $\mu(t) = \mu$, its auto-covariance function only depends on the difference between times, $\gamma(s,t) = \sigma|t-s|$ for a $\sigma\in \mathbb{R}$, and it has a finite variance function, $\mu_2(t) = \mu_2 < \infty$.
\end{Def}
For the rest of this work, the term \textit{stationary} will be used to specify a weakly stationary process. A direct consequence of the  stationarity assumption is that the previous indicator functions get simplified. Thus, given a stationary stochastic process  $X,$ its mean function, $k$-th order centered moment, for $k>1,$ and auto-covariance function are respectively,
$$\mu = E[X_{t_1}]\mbox{, } \mu_k = E[(X_{t_1} -\mu)^k] \mbox{ and } \gamma(h) = E[(X_{t_1+h}-\mu)(X_{t_1}-\mu)],$$
which are independent of $t_1\in T.$ 

Given a sample $x_1, \ldots, x_n,$  $n\in\mathbb{N},$   of equally spaced observations of $X,$ their corresponding estimators, sample mean, sample $k$-th order centered moment and sample auto-covariance, are respectively
$$\widehat{\mu} := n^{-1}\sum_{i=1}^nx_i\mbox{, } \widehat{\mu}_k := n^{-1}\sum_{i=1}^n(x_i - \widehat{\mu})^k \mbox{ and }\widehat{\gamma}(h) := n^{-1}\sum_{i = 1}^{n-h}(x_{i+h} - \widehat{\mu})(x_i - \widehat{\mu}).$$ 
A particular case in which stationarity implies strictly stationarity are Gaussian processes.
\begin{Def}
  A stochastic process $X$ is said to be a \emph{Gaussian process} if for every finite collection $\tau = \{t_1,t_2,\ldots, t_k\} \subset T,$ the joint distribution of $\{X_t\}_{t \in \tau}$has a multivariate normal distribution.
\end{Def}
A series of mean zero uncorrelated random variables with finite constant variance is known as \emph{white noise}. If additionally, it is formed of  independent and identically distributed (i.i.d) normal random variables,  it is known as \emph{Gaussian white noise}; which is a particular case of stationary Gaussian process. For the rest of the work, $X_t \sim N(\mu,\sigma^2)$ denotes that the random variable $X_t$ is normally distributed with mean $\mu$ and variance $\sigma^2$ and $\chi^2(v)$ denotes the chi square distribution with $v$ degree freedom. 

Other classes of stochastic processes can be defined using collections of white noise, for instance, the linear process.
\begin{Def}
  Let $X$ be a stochastic process. $X$ is said to be \emph{linear} if it can be written as
  $$X_t = \mu + \sum_{i\in\mathbb{Z}}\phi_i\epsilon_{t-i},$$
  where $\{\epsilon_i\}_{i\in\mathbb{Z}}$ is a collection of white noise random variables and $\{\phi_i\}_{i\in\mathbb{Z}}$ is a set of real values such that $\sum_{i\in\mathbb{Z}} |\phi_j| < \infty.$
\end{Def}
An important class of processes is the \textit{auto-regressive moving average} ($ARMA$). \cite{Box1990} introduced it for time series analysis and forecast,  becoming very well-known in the 90s and early 21st century. 
\begin{Def}
  For any non-negative integers $p,q,$ a stochastic process $X$ is an  $ARMA(p,q)$ process if it is a stationary process and 
  \begin{equation}\label{ARMA}
    X_t = \sum_{i=0}^p \phi_iX_{t-i}  +\sum_{i=0}^q \theta_i\epsilon_{t-i},
  \end{equation}
  where $\{\phi_i\}_{i=1}^p$ and $\{\theta_i\}_{i=0}^q$  are sequences of real values with $\phi_0= 0,$ $\phi_p\neq 0,$ $\theta_0=1$ and $\theta_q\neq 0$ and $\{\epsilon_{i}\}_{i\in\mathbb{Z}}$ is a collection of white noise random variables.
\end{Def}
Particular cases of  $ARMA$ processes are the auto-regressive ($AR(p) := ARMA(p,0)$) and the mean average ($MA(q) := ARMA(0,q)$) processes. Additionally, a \emph{random-walk} is a non-stationary 
process satisfying (\ref{ARMA}) with $p=1,$ $\phi_1 = 1$ and $q=0.$ Several properties of an $ARMA$ process can be extracted from its structure. For that, the $AR$ and $MA$ polynomials are introduced
$$AR:\text{ } \phi(z) = 1-\sum_{i=0}^p \phi_i z^i \text{ and } MA:\text{ } \theta(z) = \sum_{i=0}^q \theta_i z^i,$$
where $z$ is a complex number and, as before, $\phi_0= 0,$ $\phi_p\neq 0,$ $\theta_0= 1$ and $\theta_q\neq 0.$ Conditions for stationarity, order selection and process behavior are properties studied from these two polynomials.

For modeling volatility in financial data \cite{Bollerslev1986} proposed the \textit{generalized auto-regressive conditional heteroscedastic} (GARCH) class of   processes as a generalization of the \textit{auto-regressive conditional heteroscedastic} (ARCH) processes \citep{engle1982}.

\begin{Def}
  For any $p,q \in \mathbb{N}$, a stochastic process $X$ is a $GARCH(p,q)$ process if it satisfies
  $X_t = \mu + \sigma_{t}\epsilon_t$
  with
  $$\sigma_t^2 = \alpha_0 +\sum_{i=1}^p\alpha_i \epsilon_{t-i}^2 +\sum_{i=1}^q \beta_{i}\sigma^2_{t-i}.$$
  $\mu$ is the process mean,  $\sigma_0$ is a positive constant value, $\{\alpha_i\}_{i=1}^p$ and $\{\beta_i\}_{i=1}^q$ are non-negative sequences of real values and $\{\epsilon_{t}\}_{t \in T}$ is a collection of i.i.d. random variables. 
\end{Def}

A more general class of processes are the \textit{state-space models} ($SSMs$), which have gained popularity over the years because they do not impose on the process common restrictions such as linearity or stationarity and are flexible in incorporating the process different characteristics  \citep{OBrien2010}.
They are widely used 
for smoothing \citep{west2006} and forecasting \citep{Rob2007} in time series analysis.
The main idea is to model the process dependency with two equations: the \textit{state equation}, which models how parameters change over time and the \textit{innovation equation}, which models the process in terms of the parameters. 
Some particular SSMs that analyze the level, trend and seasonal components of the process are known as \textit{error, trend, and seasonal} (ETS)  models. There are over 32 different variations of  ETS models \citep{Hyndman2008}. 
One of them is the \textit{multiplicative error, additive trend-seasonality} $(ETS(M,A,A))$ model.

\begin{Def}\label{ETS}
A SSM process $X$ follows an ETS(M,A,A) model, if the process accepts  
	$$X_t = [L_{t-1} +T_{t-1} + S_{t-1}](1 + \epsilon_t)$$ as  innovation equation and
	\begin{eqnarray*}L_t &= &L_{t-1} +T_{t-1} +\alpha (L_{t-1} +T_{t-1} +S_{t-m})\epsilon_t\\
	T_t &= &T_{t-1} + \beta (L_{t-1} +T_{t-1} +S_{t-m})\epsilon_t\\
	S_t &= &S_{t-m} + \gamma (L_{t-1} +T_{t-1} +S_{t-m})\epsilon_t,\end{eqnarray*}  
	as 	state equations.
$\alpha, \beta,\gamma \in [0,1]$, $m\in\mathbb{N}$ denotes the period of the series and $\{\epsilon_t\}$ are i.i.d normal random variables. For each $t\in\mathbb{Z},$ $L_t$, $T_t$ and $S_t$ represent respectively the level, trend and seasonal component. 
\end{Def}
%
%
\addcontentsline{toc}{section}{Normality tests for stationary processes}
\section{Normality tests for stationary processes} \label{sec:normality}
 Extensive literature exists on goodness of fit tests for normality under the assumption of independent and  identical distributed random variables, including Pearson's chi-squared test \citep{Pearson1895}, Kolmogorov-Smirnov  test \citep{Smirnov1948},  Anderson-Darling  test \citep{anderson1952}, SK test \citep{jarque1980} and  Shapiro-Wilk test \citep{	SWtest1965,Royston1982} among others. These procedures have been widely used in many studies and applications, see \cite{Dagostino1987} for further details. There are no results, however, showing that the above tests are consistent in the context of stationary processes, case in which the independence assumption is violated. For instance, \cite{Gasser1975} provides a simulation study where Pearson's chi-squared test has an excessive rejection rate under the null hypothesis for dependent data. For this matter, several tests have been proposed over the years; a selection of which we reference here. \cite{epps1987} provides a test based on the characteristic function and a similar test is proposed by \cite{Hinich1982} based on the process' spectral density function  \citep[for further insight]{Berg2010}. \cite{Gasser1975} gives a correction of the SK test, with several modifications made in \cite{Lobato2004,bai2005,MarianZach2017}, which are popular in many financial applications. \cite{Meddahi2005} constructs a test based on Stein's characterization of a Gaussian distribution. Using the random projection method \citep{Cuesta2007}, \cite{nietoreyes2014} build a test that upgrades the performance of \cite{epps1987} and \cite{Lobato2004} procedures. Furthermore, \cite{vavra2017} proposed a bootstrap approximation of the \cite{anderson1952} test statistic for stationary linear processes.

Despite the existing literature, consistent implementations of goodness of fit test for normality of stationary processes in programming languages such as \proglang{R} or \proglang{Python} are limited. We present here the \pkg{nortsTest} package: it performs the tests proposed in \cite{epps1987}, \cite{Lobato2004}, \cite{nietoreyes2014} and \cite{vavra2017}. To install the latest release version of \pkg{nortsTest} from \code{CRAN}, type \code{install.packages("nortsTest")} 
within \proglang{R}. The current development version can be installed from \proglang{GitHub} using the next code:
\begin{CodeChunk}
 \begin{CodeInput}
R> if (!requireNamespace("remotes")) install.packages("remotes")		
R> remotes::install_github("asael697/nortsTest",dependencies = TRUE)
 \end{CodeInput}
\end{CodeChunk}
Additionally, the package offers visualization functions for descriptive time series analysis and several diagnostic methods for checking stationarity and normality assumptions for the most used time series models of several \proglang{R} packages. To elaborate on this, Subsection \ref{sub:Software} introduces the package functionality and software and Subsection \ref{sub:uroot} provides an overview of the used methods for checking stationarity and seasonality. Finally, Subsections \ref{sub:epps}-\ref{sub:vavra} present a general framework of each of the implemented test and their functionality  by providing simulated data examples.
\addcontentsline{toc}{subsection}{Software}
\subsection{Software}\label{sub:Software}
The package works as an extension of the \pkg{nortest} package \citep{nortest2015}, which performs normality tests in random samples but for independent data. The  building block  functions of the \pkg{nortsTest} package are:
\begin{itemize}
	\item \code{epps.test()}, function that implements the  test of Epps,
	\item \code{lobato.test()}, function that implements the  test of Lobato and Velasco,
	\item \code{rp.test()}, function that implements the random projection test of Nieto-Reyes, Cuesta-Albertos and Gamboa, and
	\item \code{vavra.test()}, function that implements the test of Psaradaki and Vavra.
\end{itemize}
Each of these functions accepts a \code{numeric} (\textit{numeric}) or \code{ts} (\textit{time series}) class object for storing data, and returns a \code{htest} (\textit{hypothesis test}) class object with the main results for the test. To guarantee the accuracy of the results, each test performs unit root tests for checking stationarity and seasonality (see Subsection \ref{sub:uroot}) and displays a warning message if any of them not satisfied. 

For visual diagnostic, the package offers different plot functions based on the \pkg{ggplot2} package \citep{ggplot2}: the \code{autoplot()} function plots \code{numeric}, \code{ts} and \code{mts} (\textit{multivariate time series}) classes while the \code{gghist()} and \code{ggnorm()} functions are for plotting histogram and qq-plots respectively; and on the  \pkg{forecast} package \citep{Rob2007}: \code{ggacf()} and \code{ggPacf()}  for the display of the  auto-correlation and partial auto-correlations functions respectively.

Furthermore, inspired in the function \code{check.residuals()} of the \pkg{forecast} package, we provide the \code{check_residuals()} function for checking assumptions of the model using the estimated residuals. Thus this function checks stationarity, seasonality (\textit{see Subsection \ref{sub:uroot}}) and normality, presenting a report of the used tests and conclusions. If the \code{plot} option is \code{TRUE}, the function displays several plots for visual checking. An illustration of these functions is provided in Subsection \ref{sub:application}, where we show the functions details and their utility for assumptions commonly checked in time series modeling.
\addcontentsline{toc}{subsection}{Test for stationarity}
\subsection{Tests for stationary}\label{sub:uroot}
For checking stationarity, the \pkg{nortsTest} package uses \textit{unit root} and \textit{seasonal unit-roots}  tests. These tests work similarly, checking whether a specific process follows a random-walk model, which clearly is a non-stationary process. 
\addcontentsline{toc}{subsubsection}{Unit root tests}
\subsubsection{Unit root tests}
A stochastic process $X$ is non stationary if it follows a random-walk model. This statement is equivalent to say that the AR(1) polynomial ($\phi(z) = 1 - \phi z$) of $X$ has a unit root \footnote{If $\phi = 1$, then $\phi(z) = (1 -z)$  which its only root is one}.
The most commonly used tests for unit root testing are  \textit{Augmented Dickey Fuller} \citep{dickey1984},  \textit{Phillips-Perron} \citep{Perron1988},  \textit{kpps} \citep{KppsI1992} and  \textit{Ljung-Box} \cite{Box}. The \code{urrot.test()} and \code{check_residual()} functions  perform  these tests, making use of  the  \pkg{tseries} package \citep{tseries}.
\addcontentsline{toc}{subsubsection}{Seasonal unit root tests}
\subsubsection{Seasonal unit root tests}
Let $X$ be a stationary process and $m$ be its  period\footnote{For observed data, $m$ is the number of observations per unit of time}. $X$ follows a seasonal random walk if it can be written as
$$X_t = X_{t-m} + \epsilon_t,$$
where $\epsilon_t$ is a collection of i.i.d random variables. In a similar way, the process $X$ is non-stationary if it follows a seasonal random-walk. Or equivalently, $X$ is non-stationary if the seasonal AR(1) polynomial ($\phi_m(z) = 1 - \phi z^m$) has a unit root. The \code{seasonal()} and \code{check_residuals()} functions perform the \textit{OCSB test} \citep{ocsb1988} from the \pkg{forecast} package, and the \textit{HEGY} \citep{Hegy1993} and \textit{Ch} \citep{ch1995} tests from the \pkg{uroot} package \citep{uroot}.
\addcontentsline{toc}{subsection}{The Epps' test}
\subsection{Test of Epps}\label{sub:epps}
The $\chi^2$ test for normality proposed by \cite{epps1987}  compares the empirical characteristic function of the one-dimensional marginal of the process with the one of a normally distributed random variable evaluated at certain points on the real line. Several authors, such as \cite{Lobato2004} and \cite{vavra2017}, point out that the greatest challenge in this test is its implementation procedure. 

Let $X$ be a stationary stochastic process  that satisfies
\begin{equation}\label{a}
  \sum_{t=-\infty}^{\infty}|t|^k|\gamma(t)| <\infty \mbox{ for some } k >0.
\end{equation}
The null hypothesis is that the one-dimensional marginal distribution of $X$ is a Gaussian process. As wee see in what follows, the procedure for constructing the test consists of defining a function $g$, estimating its inverse spectral matrix function, minimizing the generated quadratic function in terms of the unknown parameters of the random variable and, finally, obtaining the test statistic, which converges in distribution  to a $\chi^2.$ 

Given $N \in\mathbb{N}$ with $N > 2,$  let 
$$\Lambda :=\{\lambda:=(\lambda_1, \ldots, \lambda_N) \in \mathbb{R}^N: \lambda_i \leq \lambda_{i+1} \text{ and } \lambda_i > 0, \text{ for }  i = 1,2,\ldots,N \}$$ 
and $g:\mathbb{R}\times \Lambda \rightarrow \mathbb{R}^n$ be  a measurable function, where 
$$g(x,\lambda):= [\cos(\lambda_1x),\sin(\lambda_1x),\ldots,\cos(\lambda_Nx),\sin(\lambda_Nx)].$$
Additionally, let  $g_\theta:
\Lambda \rightarrow \mathbb{R}^N$ be a function defined by
$$g_\theta(\lambda) := \left[\mbox{Re}(\Phi_\theta(\lambda_1)),\mbox{Im}(\Phi_\theta(\lambda_1)),\ldots,\mbox{Re}(\Phi_\theta(\lambda_N)),\mbox{Im}(\Phi_\theta(\lambda_N))  \right]^t,$$
where the $\mbox{Re}(\cdot)$ and $\mbox{Im}(\cdot)$  are the real and imaginary components of a complex number and $\Phi_\theta$ is the characteristic function of a normal random variable with parameters  $\theta = (\mu,\sigma^2)\in \Theta,$ an open bounded set contained in $\mathbb{R}\times \mathbb{R}^+$. 
For any $\lambda\in\Lambda,$ let us also denote
$$\widehat{g}(\lambda) := \dfrac{1}{n}\sum_{t=1}^n [\cos(\lambda_1X_t),\sin(\lambda_1X_t),\ldots,\cos(\lambda_NX_t),\sin(\lambda_NX_t)]^t.$$ 
Let $f(v;\theta,\lambda)$ be the spectral density matrix of $\{g(X_t,\lambda)\}_{t \in\mathbb{Z}}$ at a frequency $v.$
Then, for $v = 0$, it can be estimated by
$$\widehat{f}(0;\theta,\lambda) := \dfrac{1}{2\pi n}\left(\sum_{t=1}^n \widehat{G}(X_{t,0},\lambda) +2\sum_{i=1}^{\lfloor n^{2/5}\rfloor}(1 -i/\lfloor n^{2/5} \rfloor)\sum_{t=1}^{n-i}\widehat{G}(X_{t,i},\lambda) \right),$$
where $\widehat{G}(X_{t,0},\lambda) = (\widehat{g}(\lambda) -g_\theta(\lambda))(\widehat{g}(\lambda) -g_\theta(\lambda))^t$ and $\lfloor \cdot \rfloor$ denotes the floor function. The test statistic general form under $H_0$ is
$$Q_n(\lambda) := \min_{\theta \in \Theta} \left\{ Q_n(\theta,\lambda) \right\},$$
with 
$$Q_n(\theta,\lambda):=(\widehat{g}(\lambda)-g_\theta(\lambda))^tG_n^+(\lambda)(\widehat{g}(\lambda)-g_\theta(\lambda))$$ 
where $G^{+}_n$ is the generalized inverse of the spectral density matrix $2 \pi  \widehat{f}(0;\theta,\lambda)$. Let $\widehat{\theta} = \arg \min_{\theta \in \Theta} \left\{ Q_n(\theta,\lambda) \right\}$ be the argument that minimizes $Q_n(\theta,\lambda)$ such that $\widehat{\theta}$ is in a neighborhood of $\widehat{\theta}_n = (\widehat{\mu},\widehat{\gamma}(0))$. To guarantee its' existence and uniqueness, the following assumptions are required. We refer to them as assumption $(A.)$.
\begin{itemize}
\item[$(A.)$]
	Let $\theta_0$ be the true value of $\theta = (\mu,\sigma^2)$ under $H_0$, then for every $\lambda \in \Lambda$  the following conditions are satisfied.
	\begin{itemize}
		\item $f(0;\theta,\lambda)$ is positive definite.
		\item $\Phi_\theta(\lambda)$ is twice differentiable with respect to $\theta$ in a neighborhood of $\theta_0$.
		\item The matrix $D(\theta_0,\lambda) = \dfrac{\partial \Phi_\theta(\lambda)}{\partial\theta |_{\theta = \theta_0}} \in \mathbb{R}^{N\times 2}$, for $N >2$, has rank 2.
		\item The set $\Theta_0(\lambda) := \{ \theta \in \Theta:  \Phi_\theta(\lambda_i) =  \Phi_{\theta_0}(\lambda_i), i=1, \ldots,N\}$ is a finite bounded set in $\Theta$. And $\theta$ is a bounded subset $\mathbb{R}\times \mathbb{R}^+$.
		\item $f(0;\theta,\lambda) = f(0;\theta_0,\lambda)$ and $D(\theta_0,\lambda) = D(\theta_,\lambda)$ for all $\theta \in \Theta_0(\lambda)$. 
	\end{itemize}
\end{itemize}
Under these assumptions, the Epps's main result is presented as follows.
\begin{teo}[\cite{epps1987} Theorem 2.1]
  Let $X$ be a stationary Gaussian process such that (\ref{a}) and  $(A.)$ are satisfied, then $nQ_n(\lambda)\to_d \chi^2(2N - 2)$ for every $\lambda \in \Lambda$.
\end{teo}
For the current \pkg{nortsTest} version, we define  
$\Lambda := \{(1.0,1.0,2.0,2.0)/\widehat{\gamma}(0) \}$, where $\widehat{\gamma}(0)$ is the sample variance. Therefore, the implemented test statistics converges to a $\chi^2$ distribution with two degree freedom. In the next version of the package, the user will set $\Lambda$ as desired, with the current value as default. 
\begin{eje}\label{one}
  A stationary $AR(2)$ process is drawn using a beta distribution with \code{shape1 = 9} and \code{shape2 = 1} parameters, and the  implementation of the test of Epps,  \code{epps.test()}, is performed. At significance level  $\alpha = 0.05$, the null hypothesis of normality is correctly rejected. 
\begin{CodeChunk}
 \begin{CodeInput}
R> set.seed(298)
R> x = arima.sim(250,model = list(ar =c(0.5,0.2)),rand.gen = rbeta,
                 shape1 = 9,shape2 = 1)
R> epps.test(x)
 \end{CodeInput}
%
 \begin{CodeOutput}
     Epps test

data:  x
epps = 32.614, df = 2, p-value = 8.278e-08
alternative hypothesis: x does not follow a Gaussian Process
 \end{CodeOutput}
\end{CodeChunk}
\end{eje}
\addcontentsline{toc}{subsection}{The Lobato and Velasco's test}
\subsection{Test of Lobato and Velasco}\label{sub:Lobato}
\cite{Lobato2004} provides a consistent estimator for the  corrected SK test statistic\footnote{Also known as the Jarque-Bera test, \cite{jarque1980}.} for stationary processes \citep[for further insight]{Lomincki1961,Gasser1975}. On the contrary to the test of Epps, it does not require of additional parameters for the approximation of the test sample statistic. The general framework for the test is presented in what follows.

Let $X$ be a stationary stochastic process that satisfies 
\begin{equation}\label{aLV}
 \sum_{t=0}^{\infty}|\gamma(t)| <\infty.
\end{equation}
The null hypothesis is that the one-dimensional marginal distribution of $X$ is normally distributed, that is 
$$H_0: X_t \sim N(\mu,\sigma^2) \text{ for all } t \in \mathbb{R}.$$
Let $k_q(j_1,j_2,\ldots,j_{q-1})$ be the q-th order cummulant of $X_{1},X_{1+j_1},\ldots,X_{1+j_{q-1}}$. $H_0$ is fulfilled if all the marginal cummulants above the second order are zero. In practice, it is tested just for the third and fourth order marginal cummulants, equivalently, in terms of moments, the marginal distribution is normal by testing whether $\mu_3 = 0$ and $\mu_4 = 3 \mu_2^2$.  For non correlated data, the SK test compares the SK statistic against upper critical values from a $\chi^2(2)$ distribution \citep{bai2005}. For a Gaussian process $X$ satisfying (\ref{aLV}), it holds the limiting result
$$\sqrt{n} \binom{\widehat{\mu}_3}{\widehat{\mu}_4 -3\widehat{\mu}^2_2} \to_d N[0_2,\Sigma_F)],$$
where $0_2 := (0,0)^t \in \mathbb{R}^2$ and   $\Sigma_F := \mbox{diag}(6F^{(3)}, \text{ } 24F^{(4)}) \in \mathbb{R}^{2x2}$ is a diagonal matrix with $F^{(k)} := \sum_{j = -\infty}^{\infty}\gamma(j)^k$ for $k=3,4$ \citep{Gasser1975}. 

The following consistent estimator  in terms of the auto-covariance function is proposed in \cite{Lobato2004}
$$\widehat{F}^{(k)} = \sum_{t = 1-n}^{n-1}\widehat{\gamma}(t)[\widehat{\gamma}(t) +\widehat{\gamma}(n-|t|)]^{k-1},$$
to build a  \textit{generalized SK test} statistic
$$G := \dfrac{n \widehat{\mu}_3^2}{6 \widehat{F}^{(3)}} + \dfrac{n(\widehat{\mu}_4 -3\widehat{\mu}_2)^2}{24\widehat{F}^{(4)}}.$$
Similar to the SK test for non-correlated data, the $G$ statistic is compared against upper critical values from a $\chi^2(2)$ distribution. This is seen in the below result that establishes the asymptotic properties of the test statistics, so that the general test procedure can be constructed. The result requires the following assumptions, denoted by $(B.),$ for the process $X.$ 
\begin{itemize}
	\item[(B.)]
	\begin{itemize}
	\item $E[X_t^{16}] < \infty$  for $t \in T.$ 
	\item	$\sum_{j_1 = -\infty}^{\infty}\cdots \sum_{j_{q-1} = -\infty}^{\infty} |k_q(j_1,\ldots,j_{q-1})| < \infty \text{ for } q=2,3,\ldots,16.$
	\item $\sum_{j=1}^{\infty}\left(E \left[\text{ } E[(X_0-\mu)^k|B_j] -\mu_k\right]^2 \right)^{1/2} < \infty \text{  for } k = 3,4,$
	where $B_j$ denotes the $\sigma$-field generated by $X_t$, $t \leq -j.$ 
	\item $E\left[(X_0 -\mu)^k -\mu_k \right]^2 +2\sum_{j=1}^{\infty}E\left(\left[(X_0 -\mu)^k -\mu_k \right] \left[ (X_j -\mu)^k -\mu_k \right] \right) > 0$   for  $k = 3,4.$
	\end{itemize}
\end{itemize}
Note that these assumptions imply that the higher-order spectral densities up to order 16 are continuous and bounded.
\begin{teo}[\cite{Lobato2004}, Theorem 1]
  Let $X$ be a stationary process. If $X$ is Gaussian and satisfies (\ref{aLV}) then $G \to_d \chi^2(2)$, and under assumption (B.), the test statistic G diverges whenever $\mu_3 \neq 0$ or $\mu_4 \neq 3\mu_2^2.$
\end{teo} 
\begin{eje}
  A stationary $MA(3)$ process is drawn using a gamma distribution with \code{rate = 3} and \code{shape = 6} parameters and the test of \textit{Lobato and Velasco} is performed using the  function  \code{lobato.test()} of the proposed \pkg{nortstTest} package. At significance level  $\alpha = 0.05$, the null hypothesis of normality is correctly rejected. 

\begin{CodeChunk}
 \begin{CodeInput}
R> set.seed(298)
R> x = arima.sim(250,model = list(ma =c(0.2,0.3,-0.4)),rand.gen = rgamma,
                 rate = 3,shape = 6)
R> lobato.test(x)
 \end{CodeInput}
%
 \begin{CodeOutput}
     Lobato and Velasco's test
   		
data:  x
lobato = 62.294, df = 2, p-value = 2.972e-14
alternative hypothesis: x does not follow a Gaussian Process
 \end{CodeOutput}
\end{CodeChunk}
\end{eje}
\addcontentsline{toc}{subsection}{The random projection test}
\subsection{The Random Projections test}\label{sub:rp}
The previous two proposals only test for the normality of the one-dimensional marginal distribution of the process, which is  inconsistent against alternatives whose one-dimensional marginal is Gaussian. \cite{nietoreyes2014} provides a procedure to fully test normality of a stationary process using a Cramm\'er-Wold type result \citep{Cuesta2007} that  uses  random projections to differentiate among distributions. We show this result below. The result works for  separable Hilbert spaces,  however here, for its later application, we restrict it to $l^2,$ the space of square summable sequences over $\mathbb{N},$ with inner product $\langle \cdot,\cdot \rangle.$ 
\begin{teo}[\cite{Cuesta2007}, Theorem 3.6]
  Let $\eta$ be a dissipative distribution on $l^2$ and $Z$  a $l^2$-valued random element, then $Z$ is Gaussian if and only if 
  $$\eta\{h \in l^2: \langle Z,h \rangle \text{ has a Gaussian distribution}\} > 0.$$ 
\end{teo}
A  dissipative distribution \citep[Definition 2.1]{nietoreyes2014} is a generalization of the concept of absolutely continuous distribution to the  infinite dimensional space. To construct a dissipative distribution in $l^2,$ it is made use of the   Dirichlet process \citep{gelman2013}. In practice, the $h \in l^2$ is drawn with a stick-breaking process that makes use of beta distributions.

Let $X=\{X_t\}_{t\in\mathbb{Z}}$ be a stationary process. As $X$ is normally distributed if the process  $X^{(t)} := \{X_k\}_{k \leq t}$ is Gaussian for each   $t\in\mathbb{Z},$ using the result above, \cite{nietoreyes2014}  provides a procedure for testing  that $X$ is a Gaussian process by testing whether the process $Y^h = \{Y^h_t\}_{t \in \mathbb{Z}}$ is Gaussian.
\begin{equation}\label{proj}
 Y^h_t := \sum_{i=0}^\infty h_i X_{t-i} = \langle X^{ (t) },h \rangle,
\end{equation}
where  $\langle X^{(t)},h \rangle$ is a real random variable for each $t \in \mathbb{Z}$ and  $h\in l^2$. Thus, $Y^h$ is a stationary process constructed by the projection of $X^{(t)}$ on the space generated by $h.$ Therefore, $X$ is a Gaussian process if and only if the marginal distribution of $Y^{h}$ is normally distributed. Additionally, the hypothesis of the tests \textit{Lobato and Velasco} or \textit{Epps}, such as (\ref{a}), (\ref{aLV}), $(A)$ and $(B)$, imposed on $X$ are inherited by $Y^h$. Then, those tests can be applied to evaluate the normality of the marginal distribution of $Y^h$. Further conditions such as, a discussion on the specific beta parameters used  to construct the distribution from which to draw $h$, select a proper amount of combinations to establish the number of projections required to improve the method performance, have to be considered. All of these details are discussed in \cite{nietoreyes2014}.

Next, we summarize the test of random projections in practice:
\begin{enumerate}
	\item Select $k,$ the number of independent projections to be used (\textit{by default k = 64}).
	\item Half of the random elements in which to project are drawn from a dissipative distribution that makes use of a particular beta distribution (\textit{$\beta(2,7)$ by default}). Then the test of \textit{Lobato and Velasco}  is applied to the odd number of projected processes, and the \textit{Epps} test to the even.
	 \item The other half are drawn analogously but using another beta distribution (\textit{$\beta(100,1)$  by default}).  Then again the test of \textit{Lobato and Velasco} is applied to the odd number of projected process, and the \textit{Epps test} to the even.
	 \item The obtained $k$ $p$-values are combined using the false discover rate \citep{Benjamin2001}.
\end{enumerate}

The \code{rp.test()} function implements the above procedure. The user might provide optional parameters such as the number of projections \code{k}, the parameters of the first beta distribution \code{pars1} and those of the second  \code{pars2}. In the next example, the \code{rp.test} is applied to a stationary GARCH(1,1) process drawn using normal random variables.
\begin{eje}
  A stationary \code{GARCH(1,1)} process \footnote{A \code{GARCH(1,1)} process is stationary if the parameters $\alpha_1$ and $\beta_1$ satisfy the inequality $\alpha_1 + \beta_1 < 1$ \citep{Bollerslev1986}.} is drawn using standard normal distribution and the parameters $\alpha_0 = 0,$ $\alpha_1 = 0.2$ and $\beta_1 = 0.3.$ 
\begin{CodeChunk}
 \begin{CodeInput}	
R> set.seed(3466)
R> spec = garchSpec(model = list(alpha = 0.2, beta = 0.3))
R> x = ts( garchSim(spec, n = 300) )
R> rp.test(x,k=250)
 \end{CodeInput}
%	
 \begin{CodeOutput}
	k random projections test

data:  x
k = 250, lobato = 1.1885, epps = 3.1659, p-value = 0.8276
alternative hypothesis: x does not follow a Gaussian Process
 \end{CodeOutput}
\end{CodeChunk}
The \textit{random projections} test is applied to the simulated data with  \code{k = 250} as the number of projections (\textit{as recommended by the authors}). At significance level  $\alpha = 0.05,$ there is no evidence to reject null hypothesis of normality.
\end{eje}

The \code{random.projection()} function upgrades the \code{lobato.test()} and \code{epps.test()} functions for fully testing normality. This function generates the projected process $Y^h$ as in (\ref{proj}), the \code{shape1} and \code{shape2} function's arguments are the parameters of a beta distribution used to generate the stick-breaking process $h.$ And then, the \code{lobato.test()} or \code{epps.test()} functions can be applied to the resulting $Y^h$ process for fully testing.


\begin{eje}
  We use the AR(2) process simulated in Example \ref{one}, to fully check of normality using the \code{epps.test()} and \code{random.projection()} functions, where \code{shape1 = 100} and \code{shape2 = 1} are the arguments for generating the new projected process $Y^h$. At significance level  $\alpha = 0.05$, the null hypothesis of normality is again correctly rejected. 
\begin{CodeChunk}
 \begin{CodeInput}
R> set.seed(298)
R> x = arima.sim(250,model = list(ar =c(0.5,0.2)),rand.gen = rbeta,
                 shape1 = 9,shape2 = 1)
R> y = random.projection(x,shape1 = 100, shape2 = 1,seed = 298)		           
R> epps.test(y)
 \end{CodeInput}
%
 \begin{CodeOutput}
      Epps test
   		
data:  x
epps = 11.645, df = 2, p-value = 0.002961
alternative hypothesis: x does not follow a Gaussian Process
 \end{CodeOutput}
\end{CodeChunk}
\end{eje}
\addcontentsline{toc}{subsection}{The Psaradakis and Vavra's test}
\subsection{The Psaradakis and Vavra's test}\label{sub:vavra}
\cite{vavra2017} proposed a distance test for normality of the one-dimensional marginal distribution of a stationary process. The test is based on the \cite{anderson1952} test statistic and makes use of an auto-regressive sieve bootstrap approximation to the null  distribution of the sample test statistic. Although the test is said to be applicable to a wider class of non-stationary processes, by transforming them into stationary by means of a fractional difference operator, no theoretic result was apparently provided to sustain this transformation. Therefore, here we restrict the presentation  and implementation of the procedure to stationary processes. 

Let $X$ be a stationary process satisfying
\begin{equation}\label{aPV}
 X_t = \sum_{i=0}^{\infty}\theta_i \epsilon_{t-i} + \mu_0, \text{ } t \in \mathbb{Z},
\end{equation}
where $\mu_0 \in \mathbb{R}$, $\{\theta_i\}_{i=0}^\infty\in l^2$  with $\theta_0 = 1$ and $ \{\epsilon_t\}_{i=0}^\infty$ a collection of mean zero i.i.d random variables. The null hypothesis is that the one-dimensional marginal distribution of $X$ is normally distributed,
$$H_0: F(\mu_0 +\sqrt{\gamma(0)}x)-F_N(x) = 0, \text{ for all } x\in \mathbb{R},$$
where F is the cumulative distribution function of $X_0$, and $F_N$ denotes the standard normal cumulative distribution function. Note that  if $\epsilon_0$  is normally distributed, then the null hypothesis is satisfied. Conversely, if the null hypothesis is satisfied, then $\epsilon_0$ is normally distributed and consequently $X_0$.  

The considered test for $H_0$ is based on the Anderson-Darling distance statistic 
\begin{equation}\label{aPV1}
 A_d = \int_{-\infty}^{\infty}\dfrac{[{F_n}(\widehat{\mu}+\sqrt{\widehat{\gamma}(0)}x)-F_N(x)]^2}{F_N(x)[1-F_N(x)]}dF_N(x),
\end{equation}
where ${F_n}(\cdot)$ is the empirical distribution function associated to $F$ based on a simple random sample of size $n.$ \cite{vavra2017} propose an auto-regressive sieve bootstrap procedure to approximate the sampling properties of $A_d$ 
arguing that making use of classical asymptotic inference for $A_d$ is problematic and involved. This scheme is motivated by the fact that under some assumptions for $X,$ including (\ref{aPV}),  $\epsilon_t$ admits the representation,
\begin{equation}\label{ePV}
 \epsilon_t = \sum_{i=1}^{\infty}\phi_i(X_{t-i} - \mu_0), \text{ } t\in \mathbb{Z},
\end{equation}
for certain type of $\{\phi_i\}_{i=1}^\infty\in l^2$. The main idea behind this approach is to generate a bootstrap sample $\epsilon_t^*$  to approximate $\epsilon_t$ with a finite-order auto-regressive model. This is because the distribution of the processes $\epsilon_t$  and $\epsilon_t^*$ coincide asymptotically if the order of the auto-regressive approximation grows simultaneously with $n$ at an appropriate rate \citep{Buhlmann1997}. The procedure  makes use of the $\epsilon_t^{*'s}$ to obtain the $X_t^{*'s}$ through the bootstrap analog of  (\ref{ePV}). Then, a bootstrap sample of the $A_d$ statistic, $A_d^{*},$ is  generated making use of the bootstrap analog of (\ref{aPV}).

This test is implemented in the \code{vavra.test()} function.  1,000 sieve-bootstrap replications are used by default. The presented values are Monte-Carlo estimates of the $A_d$ statistic and \code{p.value}. 

\begin{eje}
  A stationary $ARMA$(1,1) process is simulated using a standard normal distribution, and the  implementation of the test of \textit{Psaradakis and V\'avra} is performed. At significance level $\alpha = 0.05$, there is no evidence to reject the null hypothesis of normality. 
\begin{CodeChunk}
 \begin{CodeInput}
R> set.seed(298)
R> x = arima.sim(250,model = list(ar = 0.2, ma = 0.34))
R> vavra.test(x)
 \end{CodeInput}
%	
 \begin{CodeOutput}
     Psaradakis-Vavra test
   	
data:  x
bootstrap A = 1.5798, p-value = 0.796
alternative hypothesis: x does not follow a Gaussian Process
 \end{CodeOutput}
\end{CodeChunk}
\end{eje}
\addcontentsline{toc}{section}{Simulations and data analysis}
\section{Simulations and data analysis} \label{sec:simulation}
\addcontentsline{toc}{subsection}{Numerical experiments}
\subsection{Numerical experiments} \label{sub:numerical}
Inspired in \cite{vavra2017} and \cite{nietoreyes2014}  simulation studies, this work proposes a similar procedure. This study involves drawing data from the $AR(1)$ process
\begin{equation}\label{eqAR}
 X_t = \phi X_{t-1} + \epsilon_t, \text{ }t\in\mathbb{Z}, \text{ for } \phi \in \{ 0,\pm 0.25,\pm 0.4\}
\end{equation}
where the $\{\epsilon_t\}_{t\in\mathbb{Z}}$ are i.i.d random variables. For the distribution of the $\epsilon_t$ we consider different scenarios: standard normal ($N$), standard log-normal ($logN$), Student t with 3 degrees of freedom ($t_3$), chi-squared with 10 degrees of freedom ($\chi^2(10)$) and  beta with parameters $(7, 1).$ As in \cite{vavra2017},  $m=1,000$ independent draws of the above process are generated for each pair of parameter $\phi$ and distribution. Each is taken of length $past+n,$ with $past=500$ and $n \in \{100,250,500,1000 \}$. The first 500 data points of each realization are then discarded in order to eliminate start-up effects. The $n$ remaining data points are used to compute the value of the test statistic of interest. In each particular scenario, the rejection rate is obtained by computing the proportion of times that the test is rejected among the $m$ trials.

\begin{table}[ht]
\centering
\resizebox{\linewidth}{!}{
\begin{tabular}[t]{lcccccccccc}
  \toprule
  \multicolumn{1}{c}{ } & \multicolumn{5}{c}{n = 100} & \multicolumn{5}{c}{n = 250} \\
  \cmidrule(l{3pt}r{3pt}){2-6} \cmidrule(l{3pt}r{3pt}){7-11}
  \textbf{distribution} & $\phi$ -0.4 & -0.25 & 0.0 & 0.25 & 0.4 & -0.4 & -0.25 & 0.0 & 0.25 & 0.4\\
  \midrule
  \addlinespace[0.3em]
  \multicolumn{11}{l}{\textbf{Lobato and Velasco}}\\
  \hspace{1em}$N$ & 0.045 & 0.042 & 0.039 & 0.052 & 0.037 & 0.044 & 0.054 & 0.056 & 0.050 & 0.048\\
  \hspace{1em}$logN$ & 1.000 & 1.000 & 1.000 & 1.000 & 1.000 & 1.000 & 1.000 & 1.000 & 1.000 & 1.000\\
  \hspace{1em}$t_3$ & 0.806 & 0.872 & 0.873 & 0.882 & 0.796 & 0.980 & 0.992 & 0.999 & 0.993 & 0.983\\
  \hspace{1em}$\chi^2(10)$ & 0.553 & 0.685 & 0.776 & 0.667 & 0.559 & 0.968 & 0.995 & 0.997 & 0.996 & 0.964\\
  \hspace{1em}$beta(7,1)$ & 0.962 & 0.995 & 0.999 & 0.996 & 0.958 & 1.000 & 1.000 & 1.000 & 1.000 & 1.000\\
  \addlinespace[0.3em]
  \multicolumn{11}{l}{\textbf{Epps}}\\
  \hspace{1em}$N$ & 0.066 & 0.077 & 0.084 & 0.075 & 0.070 & 0.058 & 0.059 & 0.065 & 0.073 & 0.068\\
  \hspace{1em}$logN$ & 0.825 & 0.884 & 0.969 & 0.958 & 0.948 & 0.998 & 0.999 & 1.000 & 1.000 & 1.000\\
  \hspace{1em}$t_3$ & 0.202 & 0.294 & 0.363 & 0.288 & 0.207 & 0.716 & 0.838 & 0.909 & 0.855 & 0.752\\
  \hspace{1em}$\chi^2(10)$ & 0.319 & 0.465 & 0.548 & 0.461 & 0.361 & 0.631 & 0.836 & 0.917 & 0.841 & 0.729\\
  \hspace{1em}$beta(7,1)$ & 0.781 & 0.953 & 0.991 & 0.960 & 0.887 & 0.996 & 1.000 & 1.000 & 1.000 & 0.999\\
  \addlinespace[0.3em]
  \multicolumn{11}{l}{\textbf{Random Projections, k = 10}}\\
  \hspace{1em}$N$ & 0.007 & 0.009 & 0.007 & 0.006 & 0.006 & 0.021 & 0.027 & 0.025 & 0.021 & 0.018\\
  \hspace{1em}$logN$ & 0.891 & 0.865 & 0.772 & 0.625 & 0.515 & 1.000 & 1.000 & 1.000 & 1.000 & 1.000\\
  \hspace{1em}$t_3$ & 0.293 & 0.267 & 0.204 & 0.122 & 0.098 & 0.989 & 0.993 & 0.995 & 0.983 & 0.961\\
  \hspace{1em}$\chi^2(10)$ & 0.223 & 0.231 & 0.201 & 0.131 & 0.086 & 0.954 & 0.993 & 0.992 & 0.949 & 0.840\\
  \hspace{1em}$beta(7,1)$ & 0.605 & 0.533 & 0.363 & 0.184 & 0.108 & 1.000 & 1.000 & 1.000 & 1.000 & 1.000\\
  \addlinespace[0.3em]
  \multicolumn{11}{l}{\textbf{Psaradakis and Vavra}}\\
  \hspace{1em}$N$ & 0.056 & 0.046 & 0.038 & 0.052 & 0.046 & 0.050 & 0.061 & 0.044 & 0.055 & 0.047\\
  \hspace{1em}$logN$ & 1.000 & 1.000 & 1.000 & 1.000 & 1.000 & 1.000 & 1.000 & 1.000 & 1.000 & 1.000\\
  \hspace{1em}$t_3$ & 0.714 & 0.802 & 0.850 & 0.768 & 0.644 & 0.959 & 0.987 & 0.997 & 0.990 & 0.960\\
  \hspace{1em}$\chi^2(10)$ & 0.500 & 0.692 & 0.800 & 0.660 & 0.542 & 0.911 & 0.985 & 0.995 & 0.985 & 0.922\\
  \hspace{1em}$beta(7,1)$ & 0.956 & 1.000 & 1.000 & 0.998 & 0.972 & 1.000 & 1.000 & 1.000 & 1.000 & 0.999\\
  \bottomrule
\end{tabular}
}\caption[Rejection rates estimates, part 1]{\label{tab:tab1} Rejection rate estimates over $m=1,000$ trials of the four studied goodness of fit test for the null hypothesis of normality. The data is drawn using the process defined in (\ref{eqAR}) for different values of $\phi$ and $n$ displayed in the columns and different distributions for $\epsilon_t$  in the rows. $\phi \in \{ 0,\pm 0.25,\pm 0.4\},$ $n\in\{100, 250\}$. 
}
\end{table}
\begin{table}[ht]
\centering
\resizebox{\linewidth}{!}{
  \begin{tabular}[t]{lcccccccccc}
  \toprule
  \multicolumn{1}{c}{ } & \multicolumn{5}{c}{n = 500} & \multicolumn{5}{c}{n = 1,000} \\
  \cmidrule(l{3pt}r{3pt}){2-6} \cmidrule(l{3pt}r{3pt}){7-11}
  \textbf{distribution} & $\phi$ -0.4 & -0.25 & 0.0 & 0.25 & 0.4 & -0.4 & -0.25 & 0.0 & 0.25 & 0.4\\
  \midrule
  \addlinespace[0.3em]
  \multicolumn{11}{l}{\textbf{Lobato and Velasco}}\\
  \hspace{1em}$N$ & 0.042 & 0.053 & 0.037 & 0.041 & 0.043 & 0.049 & 0.051 & 0.046 & 0.043 & 0.047\\
  \hspace{1em}$logN$ & 1.000 & 1.000 & 1.000 & 1.000 & 1.000 & 1.000 & 1.000 & 1.000 & 1.000 & 1.000\\
  \hspace{1em}$t_3$ & 0.968 & 0.987 & 0.989 & 0.983 & 0.962 & 1.000 & 1.000 & 1.000 & 1.000 & 1.000\\
  \hspace{1em}$\chi^2(10)$ & 0.902 & 0.965 & 0.996 & 0.976 & 0.880 & 1.000 & 1.000 & 1.000 & 1.000 & 1.000\\
  \hspace{1em}$beta(7,1)$ & 1.000 & 1.000 & 1.000 & 1.000 & 1.000 & 1.000 & 1.000 & 1.000 & 1.000 & 1.000\\
  \addlinespace[0.3em]
  \multicolumn{11}{l}{\textbf{Epps}}\\
  \hspace{1em}N & 0.063 & 0.077 & 0.078 & 0.072 & 0.073 & 0.051 & 0.048 & 0.052 & 0.056 & 0.062\\
  \hspace{1em}logN & 0.989 & 1.000 & 1.000 & 1.000 & 0.999 & 1.000 & 1.000 & 1.000 & 1.000 & 1.000\\
  \hspace{1em}t3 & 0.569 & 0.705 & 0.781 & 0.694 & 0.587 & 0.999 & 1.000 & 1.000 & 1.000 & 0.999\\ 
  \hspace{1em}chisq10 & 0.534 & 0.745 & 0.859 & 0.740 & 0.611 & 0.999 & 1.000 & 1.000 & 1.000 & 1.000\\
  \hspace{1em}beta(7,1) & 0.983 & 0.998 & 1.000 & 1.000 & 0.989 & 1.000 & 1.000 & 1.000 & 1.000 & 1.000\\
  \addlinespace[0.3em]
  \multicolumn{11}{l}{\textbf{Random Projections k = 10}}\\
  \hspace{1em}N & 0.016 & 0.015 & 0.012 & 0.019 & 0.017 & 0.015 & 0.016 & 0.019 & 0.018 & 0.018\\
  \hspace{1em}logN & 1.000 & 1.000 & 1.000 & 1.000 & 1.000 & 1.000 & 1.000 & 1.000 & 1.000 & 1.000\\
  \hspace{1em}t3 & 1.000 & 1.000 & 1.000 & 1.000 & 0.999 & 1.000 & 1.000 & 1.000 & 1.000 & 1.000\\
  \hspace{1em}chisq10 & 1.000 & 1.000 & 1.000 & 1.000 & 0.993 & 1.000 & 1.000 & 1.000 & 1.000 & 1.000\\
  \hspace{1em}beta(7,1) & 1.000 & 1.000 & 1.000 & 1.000 & 1.000 & 1.000 & 1.000 & 1.000 & 1.000 & 1.000\\
  \addlinespace[0.3em]
  \multicolumn{11}{l}{\textbf{Psaradakis and Vavra}}\\
  \hspace{1em}N & 0.064 & 0.046 & 0.048 & 0.038 & 0.050 & 0.055 & 0.049 & 0.045 & 0.057 & 0.042\\
  \hspace{1em}logN & 1.000 & 1.000 & 1.000 & 1.000 & 0.998 & 1.000 & 1.000 & 1.000 & 1.000 & 1.000\\
  \hspace{1em}t3 & 0.908 & 0.972 & 0.982 & 0.958 & 0.896 & 1.000 & 1.000 & 1.000 & 1.000 & 1.000\\
  \hspace{1em}chisq10 & 0.824 & 0.954 & 0.988 & 0.958 & 0.856 & 1.000 & 1.000 & 1.000 & 1.000 & 1.000\\
  \hspace{1em}beta(7,1) & 1.000 & 1.000 & 1.000 & 0.998 & 1.000 & 1.000 & 1.000 & 1.000 & 1.000 & 1.000\\
  \bottomrule
  \end{tabular}
}\caption[Rejection rates estimates, part 2]{\label{tab:tab2} Rejection rate estimates over $m=1,000$ trials of the four studied goodness of fit test for the null hypothesis of normality. The data is drawn using the process defined in (\ref{eqAR}) for different values of $\phi$ and $n$ displayed in the columns and different distributions for $\epsilon_t$  in the rows.  $\phi \in \{ 0,\pm 0.25,\pm 0.4\},$ $n\in\{500, 1000\}$.}
\end{table}

\textit{Tables} \ref{tab:tab1} and \ref{tab:tab2}  present the rejection rate estimates. For every process of length $n,$ the columns represent the used $AR(1)$ parameter, and the rows the distribution use to draw the process. The obtained results are consistent with those obtained in the publications where the different tests were proposed. As expected,  rejection rates are around 0.05 when the data is drawn making use of the standard normal distribution, as in this case the data is drawn from a Gaussian process. Conversely, high rejection rates are registered for the other distributions. Although  low rejection rates  are observed for instance for the $\chi^2(10)$ distribution in the cases of  the \textit{Epps} and \textit{random projection} test, they consistently tend to 1 when the length of the process, $n,$ increases. 
Furthermore, for the random projections test, the number of projections used in this study is $k = 10$, which is by far a lower number than the recommended by \cite{nietoreyes2014}. However, even in these conditions, the obtained results are satisfactory, having even better performance than the tests of \cite{epps1987}, or \cite{vavra2017}.

\addcontentsline{toc}{subsection}{Real data application}
\subsection{Real data application}\label{sub:application}
\begin{figure}[ht]
  \centering
  \includegraphics[width=.5\linewidth]{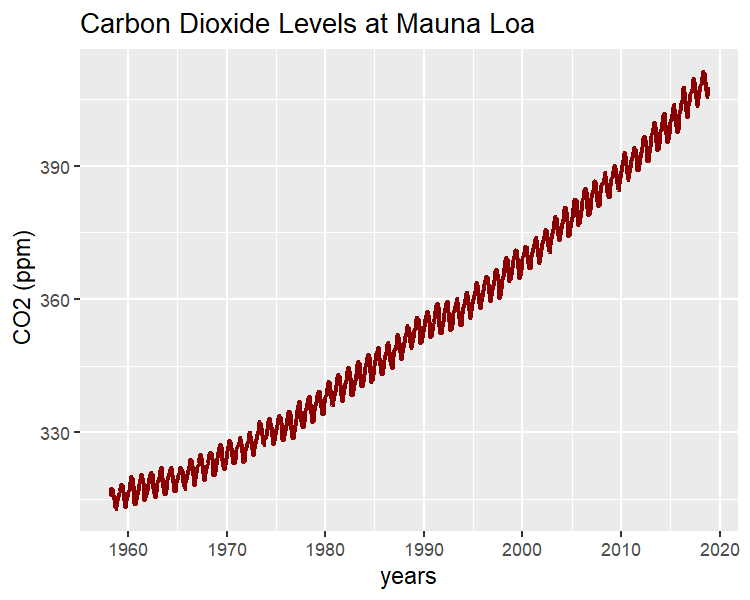}
  \includegraphics[width=.5\linewidth]{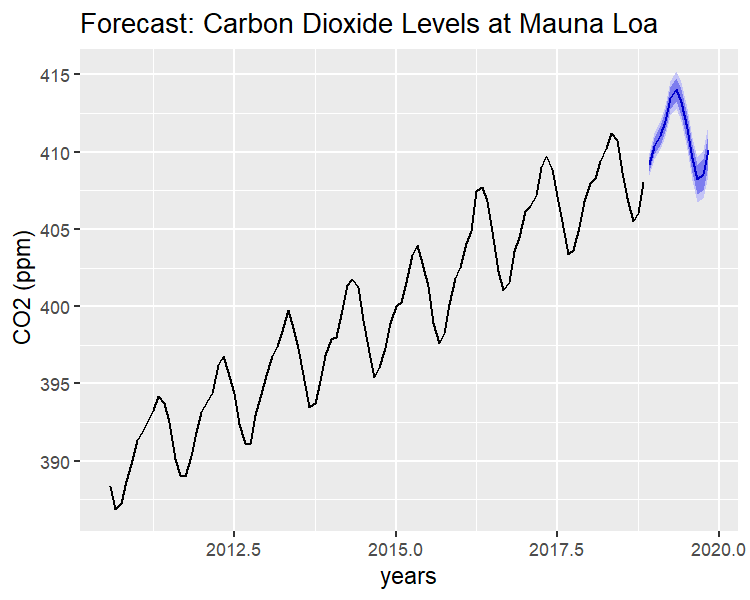}
  \caption[CO2 levels at Mauna Loa]{Left panel: CO2 Levels at Mauna Loa, time-series plot. The cardox data show a positive tendency and strong seasonality. Right panel: forecast of the next 12 months for the CO2 levels at Mauna Loa, the model's predictions capture the time-series behaviour.}
  \label{fig:fig1}
\end{figure}

As an illustrative example, we analyze the monthly mean carbon dioxide, in parts per million (\textit{ppm}), measured at the Mauna Loa Observatory, in Hawaii, from March 1958 to November 2018. 
The carbon dioxide data measured as the mole fraction in dry air on Mauna Loa constitute the longest record of direct measurements of $CO2$ in the atmosphere. This dataset is available in  the \pkg{astsa} package \citep{astsa} under the name \textit{cardox} data and it is displayed in the left panel of Figure \ref{fig:fig1}.

The objective of this subsection is to  propose a model to analyze this time series and check the assumptions on the residuals  of the model using our implemented \code{check_residuals()} function. The time series clearly has trend and seasonal components (see left panel of Figure \ref{fig:fig1}), therefore, an adequate model that filters both components has to be selected. 
We propose 
an ETS model. For its implementation, we make use the \code{ets()} function  from the \pkg{forecast} package \citep{Rob2007}. This function fits 32 different ETS models  and selects the best model according to information criterias such as  \textit{Akaike's information criteria} (AIC) or \textit{Bayesian Information criteria} (BIC) \citep{BIC2006}.
The results provided by the \code{ets()} function are:
\begin{CodeChunk}
 \begin{CodeInput}
R> library(forecast)
R> library(astsa)
R> model = ets(cardox)
R> summary(model)
 \end{CodeInput}
%	
 \begin{CodeOutput}
ETS(M,A,A) 
   
Call:
 ets(y = cardox) 
   
  Smoothing parameters:
    alpha = 0.5591 
    beta  = 0.0072 
    gamma = 0.1061 
   
  Initial states:
    l = 314.6899 
    b = 0.0696 
    s = 0.6611 0.0168 -0.8536 -1.9095 -3.0088 -2.7503
           -1.2155 0.6944 2.1365 2.7225 2.3051 1.2012
   
  sigma:  9e-04
   
    AIC      AICc      BIC 
3136.280 3137.140 3214.338 
   
Training set error measures:
                  ME   RMSE     MAE       MPE     MAPE      MASE       ACF1
Training set 0.02324 0.3120 0.24308 0.0063088 0.068840 0.1559102 0.07275949	
 \end{CodeOutput}
\end{CodeChunk}

The resulting model proposed for analyzing the \textit{carbon dioxide} data in \textit{Mauna Loa} is an $ETS[M,A,A]$ model. The parameters $\alpha, \beta \text{ and } \gamma$ (see Definition \ref{ETS}) have being estimated using the least squares method.
 If the  assumptions on the model are satisfied, then the errors of the model  behave like a Gaussian stationary process. To check it, we make use of the function \code{check_residuals()}. For more details on the compatibility of this function with the models obtained by other packages see the \pkg{nortsTest} repository.
In the following, we display the results of using the \textit{Augmented Dickey-Fuller} test  (\textit{Subsection }\ref{sub:uroot}) to check the stationary assumption and the \textit{random projection} test with \code{k = 64} projections to check the normality assumption. For the other test options see the function's documentation.
\begin{CodeChunk}
 \begin{CodeInput}
R> check_residuals(model,unit_root = "adf",normality = "rp",plot = TRUE)
 \end{CodeInput}
%
 \begin{CodeOutput}
 *************************************************** 
    	
 Unit root test for stationarity: 
   	
    Augmented Dickey-Fuller Test
   	
data:  y
Dickey-Fuller = -9.7249, Lag order = 8, p-value = 0.01
alternative hypothesis: stationary
   	
   	
 Conclusion: y is stationary
 *************************************************** 
   	
 Goodness of fit test for Gaussian Distribution: 
    
    k random projections test
   
data:  y
k = 64, lobato = 3.679, epps = 1.3818, p-value = 0.5916
alternative hypothesis: y does not follow a Gaussian Process
   
   
 Conclusion: y follows a Gaussian Process
   	
 *************************************************** 
 \end{CodeOutput}
\end{CodeChunk}

The obtained results indicate  that the null hypothesis of non-stationarity is rejected at significance level $\alpha = 0.01.$ Additionally, there is no evidence to reject the null hypothesis of normality at significance level $\alpha = 0.05.$ Consequently, we conclude that the residuals follow a stationary Gaussian process, having that the 
resulting $ETS[M,A,A]$ model adjusts well to the \textit{carbon dioxide} data in \textit{Mauna Loa}.  

\begin{figure}[ht]
  \centering
  \includegraphics[width=.7\linewidth]{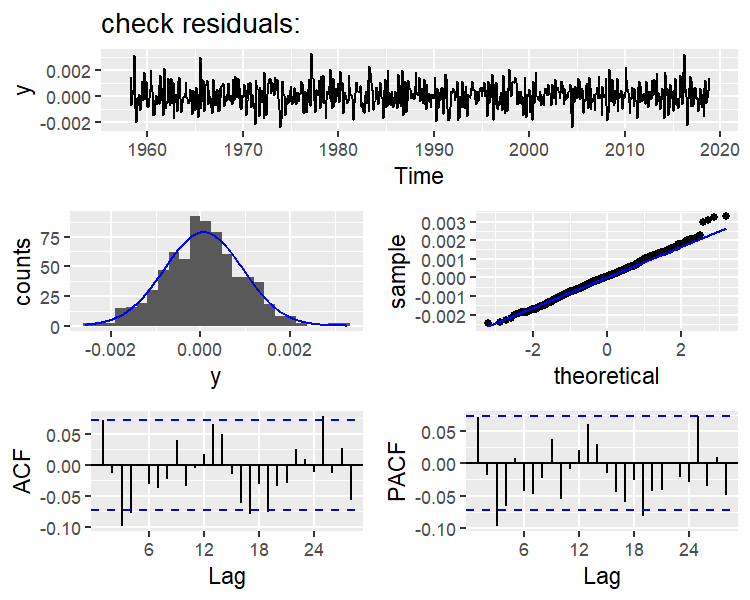}
  \caption[Check residuals plot]{Check residuals plot for the ETS(M,A,A) model. The upper panel shows the residuals time-series plot, showing small oscillations around zero, which insinuates stationarity. The middle plots are the residuals histogram (middle-left) and quantile-quantile plot (middle-right), both plots suggest that the residuals have a normal distribution. The lower panel shows the autocorrelation functions, for both plots, the autocorrelations are close to zero giving the impression of stationarity.}
  \label{fig:fig2}
\end{figure}

In the above displayed \code{check_residuals()} function, the \code{plot} argument is set to \code{TRUE}. The resulting plots are shown in Figure \ref{fig:fig2}. The plot in the  \textit{top} panel  and the auto-correlation plots in the bottom panels insinuate that the residuals have a stationary behaviour. The \textit{top} panel plot shows slight oscillations around zero and the auto-correlations functions in the \textit{bottom} panels have values close to zero in every lag. The histogram and qq-plot in the   \textit{middle} panels suggest that the marginal distribution of the residuals is normally distributed. Therefore, Figure \ref{fig:fig2} agrees with the reported results, indicating that the  assumptions of the model are satisfied. 

As the assumptions of the model have been  checked, it can be used for instance  to forecast. The result of applying the following function is displayed in the right panel of Figure \ref{fig:fig1}. It presents the Carbon dioxide data for the last 8 years and a forecast of the next 12 months. It is observable from the plot that  the model captures the process trend and periodicity. 

\begin{CodeChunk}
 \begin{CodeInput}
R> autoplot(forecast(model,h = 12),include = 100,xlab = "years",
    ylab = "CO2 (ppm)",main = "Forecast: Carbon Dioxide Levels at Mauna Loa")
 \end{CodeInput}
\end{CodeChunk}
\addcontentsline{toc}{section}{Conclusion}
\section{Conclusions} \label{sec:conclusion}
This work gives a  general overview of a careful selection of tests for normality in stationary process, which consists of the  majority of  available types of test for this matter. It additionally provides examples that illustrate each of the test implementations. 

For independent data, the \pkg{nortest} package \citep{nortest2015} provides five different tests for normality, the \pkg{mvnormtest} package \citep{mvnormtest2012} performs the Shapiro-Wilks test for multivariate data and the \pkg{MissMech} package \citep{Mortaza2014} provides tests for normality in multivariate incomplete data. To test normality of dependent data, some authors such as \cite{vavra2017,nietoreyes2014} have available undocumented \proglang{Matlab} code; mainly only useful for re-doing their simulation studies. To our knowledge, however, no consistent implementation or package of a selection of tests for normality has been done before. Therefore, the \pkg{nortsTest} is the first package that provides implementations of  tests for normality in stationary process. 

For checking model's assumptions, the \pkg{forecast} and \pkg{astsa} packages contain functions for visualization diagnostic. Following the same idea, \pkg{nortsTest} provides similar diagnostic methods; in addition to a results report of testing stationarity and normality, the main assumptions for the residuals in time series analysis. 
\addcontentsline{toc}{section}{Future work and projects}
\section*{Future work and projects}\label{future}
The second version of  the \pkg{nortsTest} package will incorporate (i) additional tests such as Bispectral \citep{Hinich1982} and Stein's characterization \citep{Meddahi2005},  (ii) upgrades in the optimization and bootstrap procedures of the tests of \textit{Epps} and \textit{Psaradaskis \& Vavra}, for faster performance, and (iii)  the creation of different implementations of the Skewness-Kurtosis test besides the one of  \textit{Lobato \& Velasco}.  Further future work will include a Bayesian version of a \textit{residuals check} procedure that makes use of the random projection method.
%
%
\section*{Acknowledgments}
I.A.M. thanks the Carolina Foundation for the fellowship that has led to this work. A.N-R. is partially supported by the Spanish Ministry of Science, Innovation and Universities grant MTM2017-86061-C2-2-P. 



\bibliography{refs}
\newpage

%
\end{document}